\documentclass[preprint]{revtex4-1}
\usepackage{graphicx}
\usepackage{amsmath}
\usepackage{bm}
\usepackage{amstext}
\usepackage{amsxtra}
\usepackage{float}
\usepackage{upgreek}

\newcommand{\To}{T_c^0}
\newcommand{\kB}{k_{\rm B}}
\newcommand{\dT}{\Delta T_c}
\newcommand{\lo}{\lambda_0}

\newcommand{\Nc}{N_{\rm c}}
\newcommand{\Ncid}{N_{\rm c}^{0}}
\newcommand{\Nt}{N}
\newcommand{\aho}{a_{\rm ho}}

\newcommand{\mumf}{\mu_c^{\rm MF}}
\newcommand{\nmf}{n_c^{\rm MF}}

\begin{document}

\title{Effects of Interactions on Bose-Einstein Condensation}

\author{Robert P. Smith}
\affiliation{Cavendish Laboratory, University of Cambridge, J. J. Thomson Avenue,
Cambridge, CB3 0HE, United Kingdom }

\begin{abstract}
Bose-Einstein condensation is a unique phase transition in that it is not driven by inter-particle interactions, but can theoretically occur in an ideal gas, purely as a consequence of quantum statistics. This chapter addresses the question  \emph{`How is this ideal Bose gas condensation modified in the presence of interactions between the particles?' }   This seemingly simple question turns out to be surprisingly difficult to answer.  Here we outline the theoretical background to this question and discuss some recent measurements on ultracold atomic Bose gases that have sought to provide some answers.
\end{abstract}

\maketitle

\section{Introduction}
\label{smith:sec:Intro}

Unlike the vast majority of phase transitions, Bose-Einstein condensation (BEC) is not driven by inter-particle interactions but can theoretically occur in an ideal (non-interacting) gas, purely as a consequence of quantum statistics. However, in reality, interactions are needed for a Bose gas to remain close to thermal equilibrium. It is thus interesting to discuss if something close to ideal gas BEC can be observed in a real system and what happens in the vicinity of the BEC transition in the presence of inter-particle interactions.   These simple questions have not been easy to answer, either theoretically or experimentally.

The theoretical foundations for studying the effect of interactions on Bose condensed systems were laid over half a century ago by Bogoliubov \cite{Bogoliubov:1947}, Penrose and Onsager \cite{Penrose:1956}, and Belieav \cite{Beliaev:1958}, among others. These works initially focused on zero-temperature properties and were extended to nonzero temperature in the pioneering papers of Lee, Huang and Yang \cite{Lee:1957b, Lee:1957a, Lee:1958, Huang:1957}. At that time the main experimental system was liquid He-4 in which the inter-particle interactions are strong, making connections with theory difficult. The realisation in 1995 of BEC in weakly interacting ultracold atomic gases \cite{Anderson:1995, Davis:1995a} thus opened up the possibility to experimentally revisit some of these long-discussed questions. This was further aided by, among other advances, the use of Feshbach resonances to tune the interaction strength in atomic gases \cite{Inouye:1998, Courteille:1998}.  Thus, in the last 20 years the study of ultracold Bose gases has been very successful. However, the fact that, until recently, ultracold atoms were confined using harmonic potentials has hindered the study of the BEC transition itself. This is due to the resulting inhomogeneous density distribution which often masks the most interesting interaction effects and also makes direct comparison with theory challenging.

In this chapter we review some recent experimental investigations using atomic Bose gases that have sought to study the role of interactions on BEC. We will mainly focus on  measurements in harmonic traps, paying particular attention to how these results relate to the physics in a homogeneous system. We also discuss some of the first measurements on a homogeneous atomic Bose gas and the possibilities that such measurements present in the future.  Note that we will focus here on three dimensional systems close to the BEC transition temperature. Some parts of the present chapter were also discussed in a previous chapter written by the author \cite{Smith:2013}.

The outline of the chapter is as follows. In section 2 we recap some theoretical background related to homogenous Bose gases in the presence of repulsive contact interactions and outline how these results may be applied to a harmonically trapped system using the local density approximation (LDA). Then in sections 3, 4 and 5 we focus on the experimental investigations of three aspects of the BEC transition in the presence of interactions:
(i) the statistical mechanism of the BEC transition,
(ii) the transition temperature, and
(iii) the critical behaviour near the transition.

\section{Theoretical background}
\label{smith:sec:Theory}

In this section we provide the key theoretical points necessary to understand and interpret the experimental results presented in the rest of the chapter. The treatment we give here is by necessity brief and we refer the interested reader to more comprehensive reviews \cite{Dalfovo:1999,Pethick:2002,Pitaevskii:2003,Andersen:2004}.
We first outline the expected behaviour for an ideal homogeneous gas before going on to consider the effect of weak interactions on such a gas. Finally, we consider how these homogeneous results may be applied to a harmonically trapped gas.

\subsection{Homogeneous ideal Bose gas}

\begin{figure}[t]
\centering
\includegraphics[width=1\columnwidth]{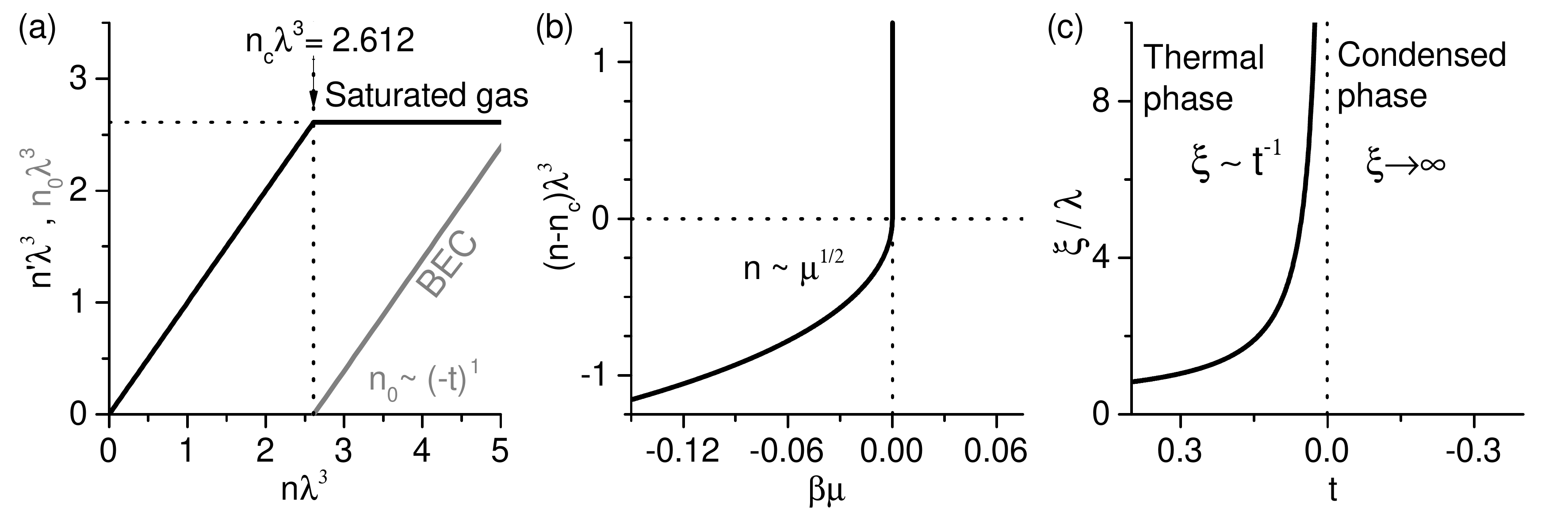}
\caption{
Ideal Bose gas condensation.
(a) Thermal $n'$ (black line) and condensed $n_0$ (grey line) density plotted versus the total density $n$, at a fixed temperature. As atoms are added  $n'=n$ and $n_0=0$ until the critical atom density $n_c$ is reached. At this point the excited states of the system saturate, and for $n> n_c$ we have $n'=n_c$ and $n_0=n-n_c\propto -t$ grows linearly.
(b) The density as a function of $\beta \mu$ close to the critical density.
(c) The correlation length diverges as $t^{-\nu}$ as we approach the transition from above with $\nu=1$; below $T_c$ the correlation length remains infinite.
}
\label{smith:fig:idealgas}
\end{figure}

In a gas of bosons of mass $m$ in equilibrium at temperature $T$ the occupation of momentum state $\bf p$ is given by the Bose distribution function,
\begin{equation}\label{smith:eq:bosep}
    f_{\bf p}=\frac{1}{\mathrm{e}^{(p^2/2m-\mu)/\kB T}-1} \; ,
\end{equation}
where $\mu$ is the chemical potential.
The total particle number $N$ can be found by summing over all possible momentum states,
\begin{equation}\label{smith:eq:sum}
    N=\sum_{p}\frac{g_{p}}{\mathrm{e}^{(p^2/2m-\mu)/\kB T}-1} \; ,
\end{equation}
where $g_{p}$ is the number of states with a given $p$. The requirement for all  terms in the sum to be real positive numbers constrains $\mu \leq p_0^2/2m$, where $p_0$ is the momentum of the ground state. As $\mu$ approaches $p_0^2/2m$ from below (which is achieved by adding particles) the ground state occupation can become arbitrarily large (see Eq.~(\ref{smith:eq:bosep})) whereas the sum of the remainder of the states (the excited states) tends to a finite number.  This helps us to understand the mechanism for Bose-Einstein condensation, namely the statistical saturation of excited states.  Saturation can be graphically  represented as shown in Fig.~\ref{smith:fig:idealgas}(a) by plotting both the ground state population and excited state population versus the total atom number for a fixed temperature. As particles are added to the system they initially populate excited states until a critical atom number is reached above which the excited state population is saturated and any additional particles must enter the ground state.

In the thermodynamic limit, in which both the volume of the system and total atom number become large, we may use the semiclassical approximation. That is, we replace the sum over excited states by an integral in order to calculate the critical atom number (or density). The excited state density $n'$, which we also call the thermal density, is given by:
\begin{equation}\label{smith:eq:densityint}
    n'=\int\frac{\mathrm{d}\mathbf{p}}{(2\pi \hbar)^3}\frac{1}{\mathrm{e}^{(p^2/2m-\mu)/\kB T}-1}=\frac{g_{3/2}(\mathrm{e}^{\mu/\kB T})}{\lambda^3} \; ,
\end{equation}
where $\lambda \; = \; [2\pi\hbar^2 /(m \kB T)]^{1/2}$ is the thermal wavelength and  $g_{3/2}(x)=\sum_{k=1}^{\infty}x^{k}/k^{3/2}$ is a polylogarithm function. Note that in this limit $p_0\rightarrow 0$ and so now $\mu \leq 0$. We can re-express this result in terms of the phase space density $\mathcal{D}=n \lambda^3$ as
\begin{equation}\label{smith:eq:psd}
    \mathcal{D'}\equiv n'\lambda^3=g_{3/2}(\mathrm{e}^{ \beta \mu}) \; ,
\end{equation}
where $\beta=1/\kB T$.
The maximum value that $\mathcal{D'}$ can take is reached when $\mu=0$ and occurs when the total density reaches a critical value $\mathcal{D}_c^0=n_c^0 \lambda^3=g_{3/2}(1)=\zeta(3/2)\approx2.612$ (where $\zeta$ is the Riemann function). Here the superscript $^0$ refers to the fact this is an ideal gas result. We can also invert this result to give the BEC transition temperature for a fixed density:
\begin{equation}\label{smith:eq:uniformTc}
   \kB T_c^0=\frac{2\pi \hbar^2}{m}\left(\frac{n}{\zeta(3/2)}\right)^{2/3} \; .
\end{equation}

In this large $N$ limit the transition to a BEC is a well defined second-order phase transition and is thus characterised by a set of critical exponents which describe how various quantities diverge when approaching the transition. Here we focus on three exponents as summarised  in Fig.~\ref{smith:fig:idealgas}.

(i) The growth of the condensate density $n_0$  below the transition temperature is described by $n_0=n-n_c \propto (-t)^{1}$ where $t$ is the \emph{reduced temperature} given by $t=(T-T_c)/T_c$.

(ii) Above $T_c$ the dependence of $n$ on $\beta \mu$ can be found by expanding Eq.~(\ref{smith:eq:psd}) for small $\beta \mu$; up to first-order this expansion (at constant volume) gives  $\mathcal{D}=\mathcal{D}_c-2\sqrt{\pi}\sqrt{-\beta \mu}$ and thus $n-n_c \propto (\mu_c-\mu)^{1/2}$ where $\mu_c$ is the critical chemical potential (for an ideal gas $\mu_c=0$). The dependence of $n$ on $\mu$ is particularly important when discussing inhomogeneous systems.

(iii)  The correlation length $\xi$  quantifies the range over which fluctuations in the gas are correlated and its divergence, described by the critical exponent $\nu$, is defined by $\xi \sim |t|^{-\nu}$. The correlation length can defined by the first-order two-point correlation function, $g_1(r) \propto \langle \hat{\Psi}^\dag(r) \hat{\Psi}(0)\rangle$,
where $\hat{\Psi}({\bf r})$ is the Bose field.  This correlation function is formally related to the Fourier transform of the momentum distribution. Far from the BEC transition ($\beta \mu \ll-1$) the Bose momentum distribution (Eq.~(\ref{smith:eq:bosep})) is approximately gaussian and thus $g_1(r)$ is short ranged and given by a gaussian of width $\lambda/\sqrt{2\pi}$. As we approach the transition long-range correlations begin to develop and for $r>\lambda$ we can approximate the correlation function as \cite{Huang:1987},
\begin{equation}
g_1(r) \propto \frac{1}{r} \exp(-r/\xi)\; .
\end{equation}
For an ideal gas $\xi/\lambda = \sqrt{1/(4 \pi \beta |\mu|)}$;  combining this result with the expansion from (ii) and the fact that for small $t$ we can write $(\mathcal{D}_c-\mathcal{D})/\mathcal{D}_c \approx\frac{3}{2}t$ gives $\xi/ \lambda =\frac{2}{3D_c}t^{-1}$. Thus, for an ideal gas BEC transition (at constant volume) $\nu=1$.

\subsection{Interacting homogeneous Bose gas}

\begin{figure}[t]
\centering
\includegraphics[width=1\columnwidth]{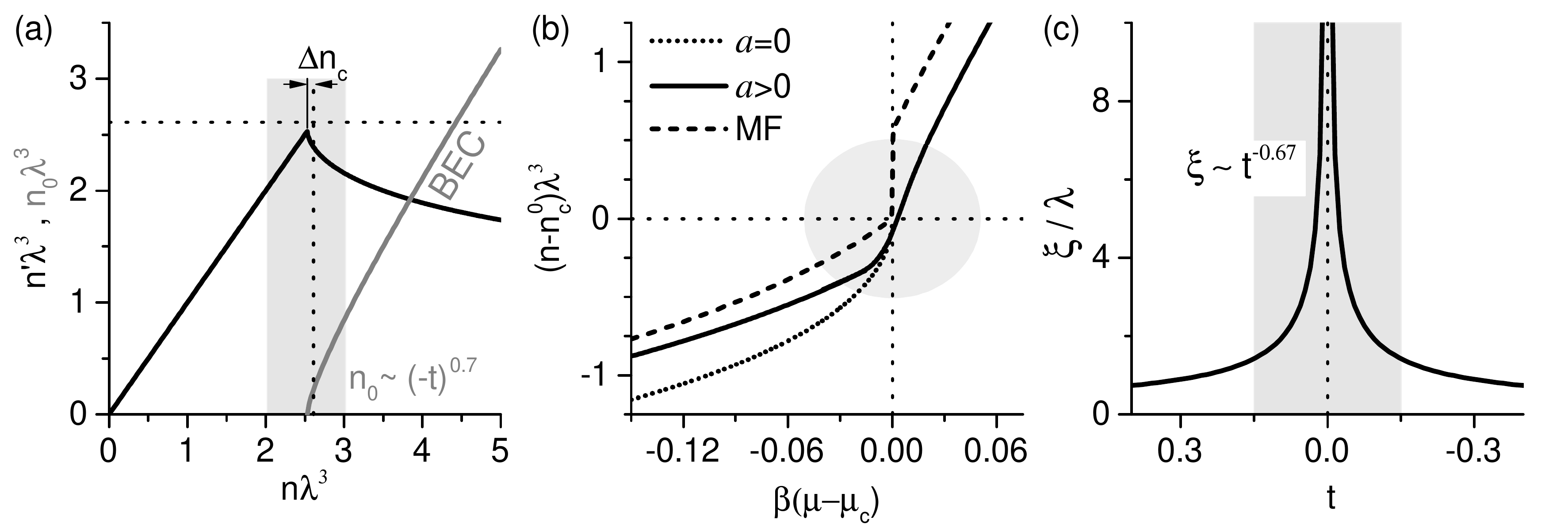}
\caption{Effects of interactions on a Bose gas. Illustrated for $a/\lambda=0.03$;  the critical region (here defined by $|t|<5 a/\lambda$) is shown in gray; the solid lines in (a) and (b) are based on Classical Field Monte-Carlo results \cite{Prokofev:2004} within the critical region and an extrapolation onto the Popov approximation outside it.
(a) Interactions both shift  the critical point and also modify the condensed and thermal densities for $n>n_c$.
(b) The dependence of $n-n_c^0$ on $(\mu-\mu_c)$  in the presence of interactions.  MF theory (dashed line) predicts $n-n_c^0 \propto \beta (\mu-\mu_c)$ and an erroneous first-order transition whereas beyond-MF theory (solid line) predicts a second order transition with an exponent between the ideal and MF values.
(c) Interactions change the correlation length critical exponent from $\nu=1$ to $\nu \simeq 0.67$.}
\label{smith:fig:interactinggas}
\end{figure}

For a dilute atomic gas the effective low-energy interaction between two atoms at $\mathbf{r}$ and $\mathbf{r}'$ can be approximated as a contact interaction $g \delta(\mathbf{r}-\mathbf{r}')$ with $g=4\pi\hbar^2 a/m$ where $a$ is the $s$-wave scattering length. Note that the dimensionless parameter which usually defines the relative strength of interactions, $a/\lambda$, is typically $\sim10^{-2}$ for ultracold atomic gases.

A simple theoretical framework in which to understand the effects of contact interactions on a Bose gas is the Hartree-Fock (HF) approximation \cite{Dalfovo:1999}. In this mean-field (MF) model one treats the thermal atoms as a ``non-interacting" gas of density $n'$ that experiences a MF interaction potential
\begin{equation}
V_{\rm int}=g(2n_0+2n')\, ,
\label{smith:eq:Vint}
\end{equation}
where $n_0$ is the condensate density.
Thus, $p^2/2m$ in Eq.~(\ref{smith:eq:bosep}) is replaced by $p^2/2m+V_{\rm int}$. Meanwhile the condensed atoms feel an interaction potential
\begin{equation}
V_{\rm int}^0=g(n_0+2n')\, ,
\label{smith:eq:Vint}
\end{equation}
where the factor of two difference in the condensate self-interaction comes about due to the lack of the exchange interaction term for particles in the same state. This HF approach does not take into account the modification of the excitation spectrum  due to the presence of the condensate, which is included in more elaborate MF theories such as those of Bogoliubov \cite{Bogoliubov:1947} and Popov \cite{Popov:1987}. However, it is often sufficient to give the correct leading order MF results. In a homogeneous system, above $T_c$, the MF interaction potential leads to a uniform energy offset which is simply compensated for by a shift in the chemical potential and thus makes no physical difference to the system. The effects of interactions on a homogeneous system at a MF-level only result from the factor of two difference in the condensate self-interaction and therefore only arise when a condensate is present (or about to appear).  All these MF theories are expected to break down as we approach $T_c$  and should only provide a good description outside the critical region ($|t| \gg a/\lambda$). Within the critical region we must revert to beyond mean-field descriptions.

Figure \ref{smith:fig:interactinggas} summarises the effect of interactions on Bose-Einstein condensation. A comparison of figures \ref{smith:fig:interactinggas} and \ref{smith:fig:idealgas} allows us to highlight several notable differences:
\begin{enumerate}

\item The gas is no longer saturated after passing through the transition but rather the excited state density decreases as we continue to increase $n$ above $n_c$. This can be understood at a mean-field level, and is due to the factor of two reduction in the condensate self-interaction which means that an atom can lower its interaction energy by an amount $g n_0$ by entering the condensate.

\item Condensation occurs at a phase space density below the ideal gas critical value of 2.612. Qualitatively this can be understood to be due to the same effect as in point 1 - that interactions favour the occupation of the condensed state. However MF-theory predicts an erroneous first-order transition and cannot predict the shift quantitatively. Theoretically calculating the shift proved notoriously difficult and took several decades for consensus to be reached (for an overview see  e.g. \cite{Arnold:2001, Baym:2001, Andersen:2004, Holzmann:2004}). It is now generally accepted that the shift is given by \cite{Arnold:2001, Kashurnikov:2001}:
    \begin{equation}
    \frac{\Delta n}{n_c}  \approx -2.7 \frac{a}{\lambda} \, ,
    \label{smith:eq:uniformncShift}
    \end{equation}
    where $\Delta n = n_c - n_c^0$. Equivalently, the $T_c$ shift at constant $n$ is
    \begin{equation}
    \frac{\Delta T_c}{T_c^0}  \approx -2/3\frac{\Delta n_c}{n_c^0} \approx 1.8 \frac{a}{\lambda} \, .
    \label{smith:eq:uniformTcShift}
    \end{equation}

\item The critical exponents of the transition are also modified. In fact, the addition of interactions results in a change of universality class to that of the so-called 3D XY-model. The beyond-MF critical exponents expected in this universality class have been calculated to high accuracy \cite{Campostrini:2006,Burovski:2006b}. Most notably the correlation length critical exponent changes from $\nu=1$ to $\nu \simeq 0.67$.
\end{enumerate}

\subsection{Bose gas in a harmonic potential}

Our approach to tackling inhomogeneous potentials is to apply the local density approximation (LDA). The effect of a potential $V(\mathbf{r})$ is to change the energy (e.g. in Eq.~(\ref{smith:eq:densityint})) from $p^2/2m$ to $p^2/2m+V(\mathbf{r})$. Within the LDA we subsume $V(\mathbf{r})$ within the chemical potential such that we have a local chemical potential,
\begin{equation}\label{smith:eq:mulocal}
\mu(\mathbf{r})=\mu-V(\mathbf{r}) \; ,
\end{equation}
and then assume that the density and momentum distribution of the gas at a point  $\mathbf{r}$ is that of a homogeneous system with chemical potential $\mu(\mathbf{r})$.

The LDA is generally valid if $V(\mathbf{r})$ is changing slowly relative to any other relevant lengthscales. For an ideal thermal gas well above $T_c$ the only relevant lengthscales are the thermal wavelength $\lambda$ and the interparticle spacing $d<\lambda$. This suggests the LDA should be good for a thermal gas as long as $\kB T\gg\hbar \omega$. However the soundness of LDA becomes less clear as $\xi$ diverges upon approaching the transition. Within LDA the critical point is reached when the maximal local $\mathcal{D}$ reaches the critical phase space density.  However, it usually makes sense to define the critical point in terms of the critical total particle number $N_c$ (as the local density is harder to measure than the total atom number in the trap).

For an ideal Bose gas in a harmonic potential, $V(\mathbf{r}) = \sum (1/2) m \omega^2_i r_i^2$, we can calculate $N_c$ by inserting  Eq.~(\ref{smith:eq:mulocal}) with this potential into Eq.~(\ref{smith:eq:psd}), setting $\mu=0$ and integrating over all space to give,
\begin{equation}\label{smith:eq:Ncid}
\Ncid=\zeta(3)\left(\frac{\kB T}{\hbar \bar{\omega}}\right)^3 \; ,
\end{equation}
where $\bar{\omega}$ is the geometric mean of the trapping frequencies and $\zeta(3)\approx1.202$. Equivalently for a fixed particle number the transition temperature is given by,
\begin{equation}\label{smith:eq:Tcid}
\kB \To=\hbar \bar{\omega} \left(\frac{N}{\zeta(3)}\right)^{1/3}.
\end{equation}

In a similar fashion to the ideal homogeneous case if we increase the total atom number $N$ at constant temperature then for $N < \Ncid$ no condensate is present and the thermal atom number  $N' = N$. However for $N > \Ncid$ the thermal component is saturated at $N' = \Ncid$. Thus for an ideal gas the basic mechanism of saturation of excited states leading to a BEC transition remains.

However, the difference between homogeneous and harmonically trapped gases is much more fundamental than a simple change in the expression for $T_c$ might suggest. Even for an ideal gas the inhomogeneous density distribution means that  as we approach the transition only the central region of the cloud is close to critical density and so the critical behaviour of the gas is quite different.    For example the divergence of $\xi$ is constrained by a short lengthscale determined by the trap.
In the presence of interactions the differences become even more manifold due to two important factors.

Firstly, unlike in a homogeneous system, when we apply LDA in a trapped system we are no longer under the constraint of constant $n$ but of constant $\mu$. This is because we are only free to vary the global chemical potential to fix the total atom number $N$ and then locally the chemical potential is set by Eq.~(\ref{smith:eq:mulocal}). Therefore the chemical potential shifts that we could dismiss in the case of a homogeneous system can now have large effects and so a trapped system can display large mean-field effects which were completely absent in the homogeneous case. In fact, as we will see, these effects often go in the opposite direction to those in a homogenous gas.

Secondly, in a trapped system only a small region (at the trap centre) is in the critical regime.  This means that the magnitude of any beyond-MF effects are significantly reduced as compared to the homogeneous case. Also, due to our first point above, the beyond-MF effects that we do see are more likely to be related to beyond-MF shifts in $\mu$ rather than the homogeneous system density shifts.

\section{Statistical mechanism of BEC in an interacting Bose gas}
\label{smith:sec:saturation}

\begin{figure}[t]
\centering
\includegraphics[width=1\columnwidth]{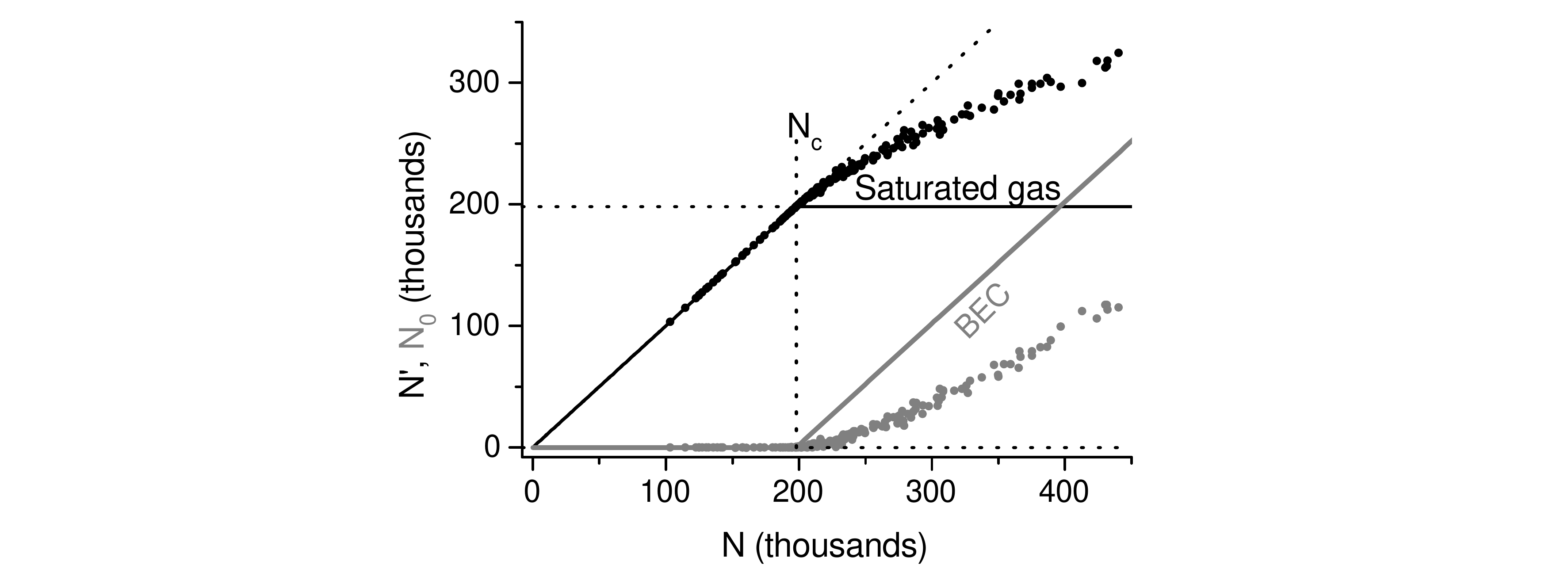}
\caption{Lack of saturation of the thermal component in a quantum degenerate Bose gas with $a/\lambda=0.01$.
Thermal atom number $N'$ (black points) and condensed atom number $N_0$ (grey points) are plotted versus the total atom number $\Nt$ for a fixed $T$.
The predictions for a saturated gas are shown by black and grey solid lines. Figure adapted from \cite{Tammuz:2011}.}
\label{smith:fig:figure1}
\end{figure}

\begin{figure}[t]
\centering
\includegraphics[width=1\columnwidth]{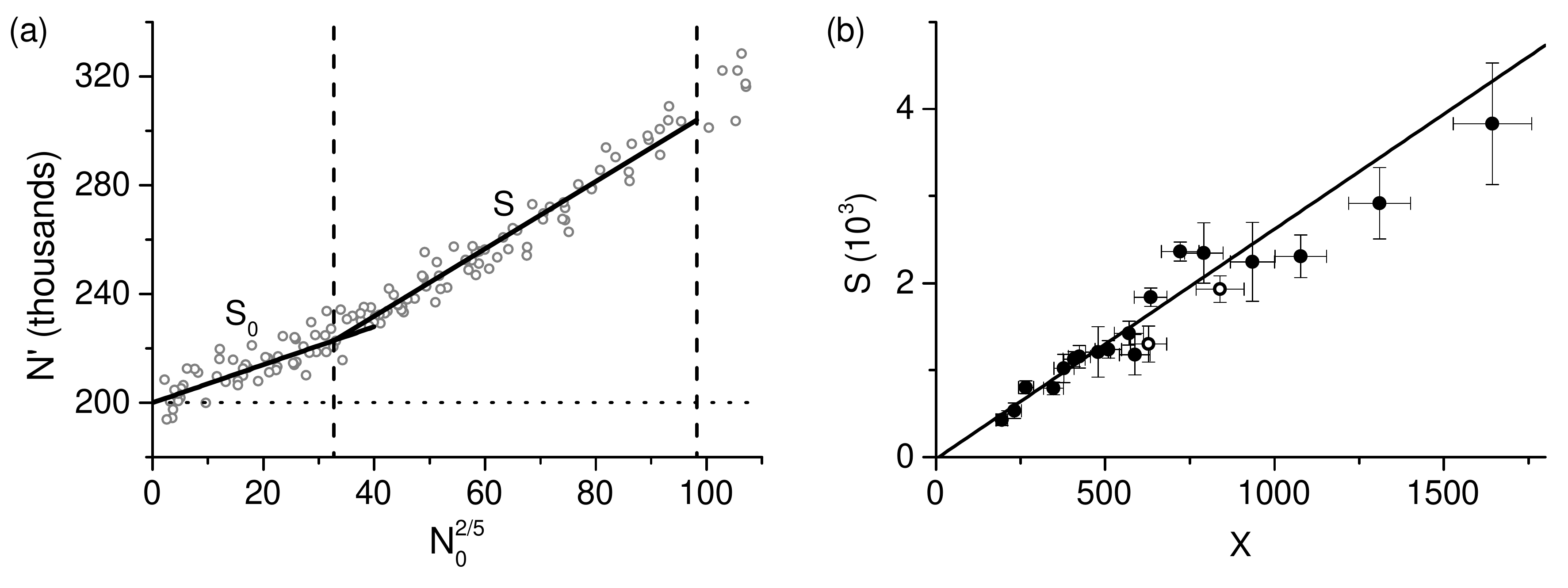}
\caption{Quantifying the lack of saturation. (a) Here $N'$ is plotted as a function of $N_0^{2/5}$ for data in Fig.~\ref{smith:fig:figure1}. The horizontal dotted line is the saturation prediction $N'=\Nc$. The two black lines show the initial slope $S_0$ and the slope $S$ for $0.1<\mu_0\, /\,\kB T<0.3$.
(b) The non-saturation slope $S$ is plotted versus the dimensionless interaction parameter $X \propto T^2 a^{2/5}$ (see text).  A linear fit (black line) gives $dS/dX = 2.6 \pm 0.3$\ and an intercept $S(0) = -20 \pm 100$, consistent with complete saturation in the ideal-gas limit.
Figures adapted from \cite{Tammuz:2011}.}
\label{smith:fig:HarmonicSat}
\end{figure}

In this section, we discuss the concept of the saturation of the excited states as the underlying mechanism driving the BEC transition, and how interactions modify this saturated-gas picture.
For superfluid $^4$He, which is conceptually associated with BEC, strong interactions preclude direct observation of purely statistical effects expected for an ideal Bose gas. This is generally thought to be in contrast to weakly interacting atomic gases in which close-to-textbook ideal BEC is expected. One might therefore expect that the saturation inequality $N' \leq \Ncid$ should be essentially satisfied in these systems. However, careful examination  \cite{Tammuz:2011} revealed that this is far from being the case for a harmonically trapped gas as shown in Fig.~\ref{smith:fig:figure1}. The drastic violation of the saturation inequality  seen in Fig.~\ref{smith:fig:figure1}, which is at first sight surprising, results from the combination of repulsive interactions and harmonic trapping. To first-order it can be explained in a simple MF picture in which we just consider the interaction of the thermal atoms with the condensate and not with other thermal atoms. This approximation works because, due to the harmonic trap, as a condensate is formed and grows, the change in the density of the condensed atoms is much larger than the change of the thermal density. Within the LDA this leads to a spatially varying interaction potential $V_{\rm int}(\mathbf{r})=2gn_0(\bf{ r})$ and results in a uniform chemical potential shift everywhere outside the condensate of,
\begin{equation}
\mu_0=gn_0 (\mathbf{r}=0)=\frac{\hbar \bar \omega}{2}
\left(
15 N_0 \frac{a}{\aho}
\right)^{2/5}\ ,
\label{smith:eq:mu0}
\end{equation}
where $N_0$ is the condensed atom number and $\aho=(\hbar/m\bar \omega)^{1/2}$ is the spatial extension of the ground state of the harmonic oscillator.
This shift in chemical potential effects the density everywhere and by integrating over the whole trap
one can predict a linear variation of $N'/N_c^0$ with the small parameter $ \beta \mu_0$:
\begin{equation}
\frac{N'}{N_c^0}=1 + \alpha  \, (\beta \mu_0) \; ,
\label{smith:eq:HF}
\end{equation}
with $\alpha = \zeta(2)/\zeta(3)\approx 1.37$. Using Eq.~(\ref{smith:eq:mu0}) we can equivalently write,
\begin{equation}
N'=N_c+S_0 N_0^{2/5} \; ,
\end{equation}
where $S_0=\alpha X$ with $X$ being the dimensionless interaction parameter:
\begin{equation}
X=\frac{\zeta(3)}{2}\left(\frac{\kB T}{\hbar \bar{\omega}}\right)^2 \left(\frac{15 \, a}{\aho}\right)^{2/5}\; .
\end{equation}

This first-order non-saturation result is identical to that obtained in more elaborate MF approximations, which only modify higher order terms.

Guided by this theory, Fig.~\ref{smith:fig:HarmonicSat}(a) shows $N'$ as a function of $N_0^{2/5}$ for the data shown in Fig.~\ref{smith:fig:figure1}. The initial growth of $N'$ with $N_0^{2/5}$ is seen to agree well with this mean-field prediction. Similar agreement of the initial slope, $\mathrm{d}N'/\mathrm{d}N_0^{2/5}\mid_{N_0\rightarrow 0}$,  with $S_0$ was observed for a wide range of interaction strength and temperature \cite{Tammuz:2011}.
Fig.~\ref{smith:fig:HarmonicSat}(a) also shows an increase of the slope for higher $N_0$ which was quantified by a course grained slope $S= \Delta[N']/\Delta [N_0^{2/5}]$ for $0.1<\mu_0\, /\,\kB T<0.3$ \cite{Tammuz:2011} and  Fig.~\ref{smith:fig:HarmonicSat}(b) summarises this non-saturation slope $S$ depends on the interaction parameter $X$.

The first and most important thing to notice is that both non-saturation slopes, $S_0$ and $S$, tend to zero for $X \rightarrow 0$.
These experiments thus confirmed the concept of a saturated Bose gas, and Bose-Einstein condensation as a purely statistical phase transition in the non-interacting limit.

They also highlighted the large effects that an inhomogeneous trapping potential can introduce in the presence of MF interactions (compare Figs.~\ref{smith:fig:interactinggas}(a) and \ref{smith:fig:figure1}). While the majority of the observed non-saturation could be explained by MF theory a significant discrepancy still remained for larger $N_0$ \cite{Smith:2013}, the origin of which is still an open question.

\begin{figure}[t]
\centering
\includegraphics[width=1\columnwidth]{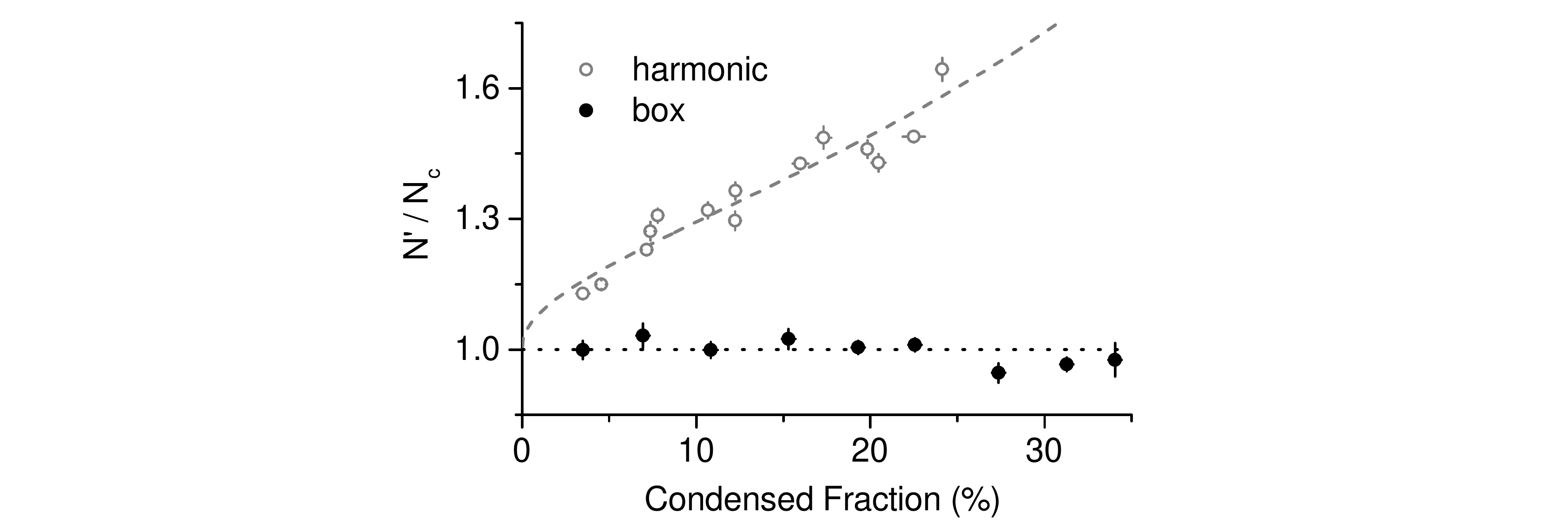}
\caption{Saturation of the thermal component in a partially condensed gas of $^{87}$Rb atoms. In the box trap the gas follows the ideal-gas prediction $N'=N_c$, whereas in the harmonic trap the thermal component is strongly non-saturated. Figure adapted from \cite{Schmidutz:2014}.}
\label{smith:fig:UniformSat}
\end{figure}

We have seen that in a harmonic trap the dominant non-saturation effect is ``geometric", arising from an interplay of the mean-field repulsion and the inhomogeneous potential. More recently, the achievement of BEC in a uniform box potential \cite{Gaunt:2013} allowed the concept of saturation to be checked for a homogeneous system where this geometric effect is absent. Figure (\ref{smith:fig:UniformSat}) directly compares the harmonically trapped and homogeneous cases and clearly shows that the saturation inequality is much more closely obeyed for a homogeneous gas. The weak interaction strength ($a/\lambda=0.006$) for these homogeneous measurements means that the expected reduction in $N'$ seen in Fig.~\ref{smith:fig:interactinggas} is not expected to be visible over this range. In the future it would be very interesting to examine the issue of saturation in more strongly interacting homogeneous gases.

\section{Transition temperature of an interacting Bose gas}
\label{smith:sec:Tc}

Having considered the effect of interactions on the saturation of the thermal component we now consider the location of the critical point itself.

\begin{figure} [t]
\centering
\includegraphics[width=1\columnwidth]{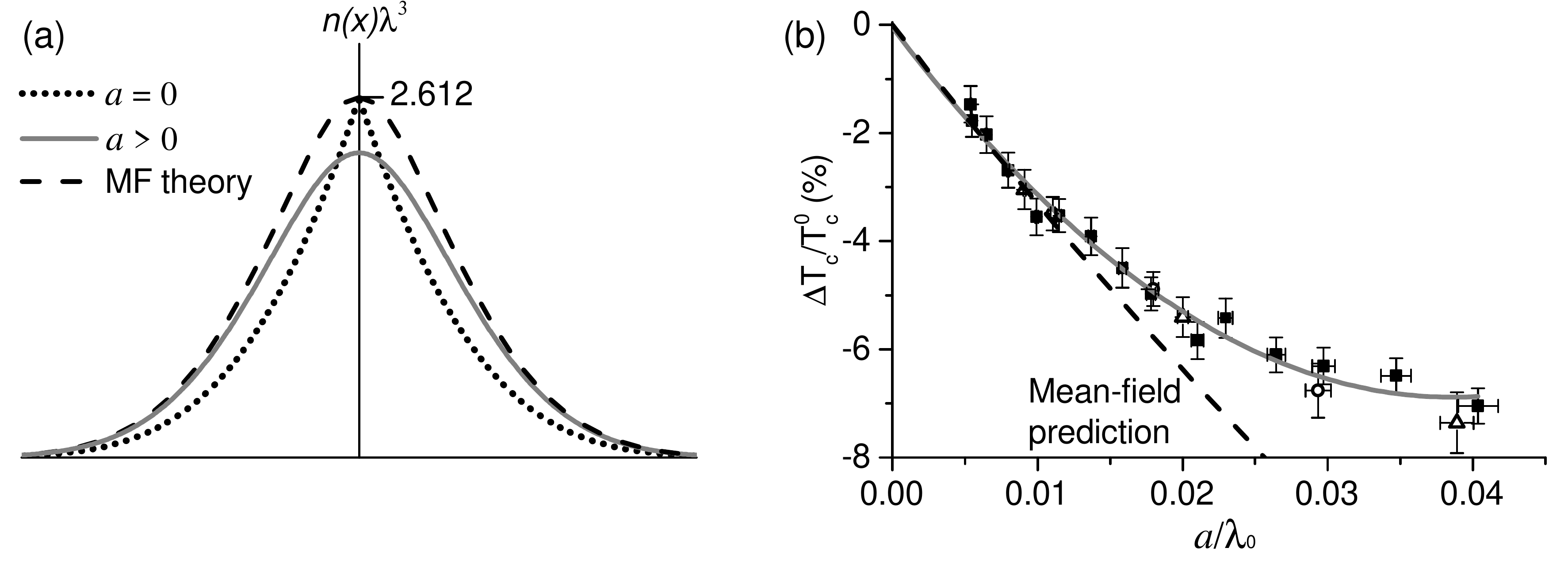}
\caption{Interaction shift of the critical temperature. (a) Opposing effects of interactions on the critical point of a Bose gas in a harmonic potential. Compared to an ideal gas (dotted line) at the same temperature, repulsive interactions reduce the critical density, but also broaden the density distribution (solid line). Mean-field theory (dashed line) captures only the latter effect, and predicts an increase of the critical atom number $N_c$ at fixed $T$, equivalent to a decrease of $T_c$ at fixed $N$. (b) Measured interaction shift of the critical temperature. The solid line shows a second-order polynomial fit to the data (see text).
Figures adapted from \cite{Smith:2011}.}
\label{smith:fig:TcShift}
\end{figure}

The problem of the $T_c$ shift in a harmonically trapped gas is even more complex than for the homogeneous case that we have already discussed. Now, as well as the expected (within LDA) reduction of the critical density which would act to increase the transition temperature
\cite{Note1}
we also have an additional mean-field geometric effect that reduces $T_c$ \cite{Giorgini:1996}.
This negative $T_c$ shift is due to the broadening of the density distribution by repulsive interactions. It arises due to the fact that while the chemical potential is shifted across the whole trap by $ V_{\rm int}(\mathbf{r}=0)=2 g n_c$, the interaction potential itself decreases with the density for ${\bf r}>0$.    To second order it can be calculated analytically using MF theory \cite{Giorgini:1996, Gaunt:2015} to give
\begin{equation}
\frac{\dT^{\rm MF}}{\To} \approx - 3.426 \, \frac{a}{\lo}+ 12.9 \, \left(\frac{a}{\lo}\right)^2 \; ,
\label{smith:eq:StringariMF}
\end{equation}
where $\Delta T_c=T_c-T_c^0$ and $\lo$ is the thermal wavelength defined at $T_c^0$.
The two opposing effects of repulsive interactions on the critical point of a trapped gas are visually summarised in Fig.~\ref{smith:fig:TcShift}(a), where we sketch the density distribution at the condensation point for an ideal and an interacting gas at the same temperature.
For weak interactions the MF effect is dominant, making the overall interaction shift $\Delta N_c(T)$ positive, or equivalently $\Delta T_c(N)$ negative.

The dominance of the negative MF shift of $T_c$ over the positive beyond-MF one goes beyond the difference in the numerical factors in Eqs.~(\ref{smith:eq:uniformTcShift}) and (\ref{smith:eq:StringariMF}). In a harmonic trap, at the condensation point only the central region of the cloud is close to criticality; this reduces the net effect of critical correlations so that they affect $T_c$ only at a higher order in $a/\lo$. The MF result of Eq.~(\ref{smith:eq:StringariMF}) should therefore be exact at first order in $a/\lo$.
The higher-order beyond-MF shift is still expected to be positive, but the theoretical consensus on its value has not been reached \cite{Houbiers:1997, Holzmann:1999b, Arnold:2001b, Davis:2006, Zobay:2009}.

Since the early days of atomic BECs there have been several measurements of the interaction $T_c$ shift in a harmonically trapped gas \cite{Ensher:1996, Gerbier:2004,Meppelink:2010}. These experiments nicely confirmed the theoretical prediction for the linear MF shift of Eq.~(\ref{smith:eq:StringariMF}), but could not discern the beyond-MF effects of critical correlations.

More recent measurements \cite{Smith:2011} shown in Fig.~\ref{smith:fig:TcShift}, which employed a Feshbach resonance, were able to discern the beyond-MF $T_c$-shift in a trapped atomic gas.
The MF prediction agrees very well with the data for $a/\lo \lesssim 0.01$, but for larger $a/\lo$ there is a clear deviation from this prediction. All the data are fitted well by a second-order polynomial
\begin{equation}
\frac{\dT}{\To} \approx b_1 \, \frac{a}{\lo} + b_2 \left( \frac{a}{\lo} \right)^2\, ,
\label{smith:eq:TrapTc}
\end{equation}
with $b_1= - 3.5 \pm 0.3$ and $b_2 = 46 \pm 5$. The value of $b_1$ is in agreement with the MF prediction of $-3.426$ whereas $b_2$ strongly excludes the MF result and its larger magnitude is consistent with the expected effect of beyond-MF critical correlations.

In order to make a connection between the experiments on trapped atomic clouds and the theory of a uniform Bose gas we also need to consider the effect of interactions on the critical chemical potential $\mu_c$.
In a uniform gas the interactions differently affect $T_c$ (or equivalently $n_c$) and $\mu_c$ at both MF and beyond-MF level. The simple MF shift $\beta \mumf = 4 \, \zeta(3/2) \, a/\lo$  has no effect on condensation. To lowest beyond-MF order we have 
\cite{Note2}:
\begin{equation}
\beta \mu_c \approx \beta \mumf + B_2 \left( \frac{a}{\lo} \right)^2 \; .
\label{smith:eq:Muc}
\end{equation}
We see that there is a qualitative difference between Eqs. (\ref{smith:eq:uniformTcShift}) and (\ref{smith:eq:Muc}). Specifically, we have $\nmf - n_c \propto a/\lo$, but $\mumf - \mu_c \propto (a/\lo)^2$.
This difference highlights the non-perturbative nature of the problem -  near criticality the equation of state does not have a regular expansion in $\mu$, otherwise one would get $\Delta n_c \propto \mu_c - \mumf$.

For a harmonic trap, within LDA the uniform-system results for $n_c$ and $\mu_c$ apply in the centre of the trap, and elsewhere the local $\mu (\mathbf{r})$ is given by Eq.~(\ref{smith:eq:mulocal}). The result for the $T_c$ shift however does not carry over easily to the non-uniform case.
Examination of Eqs.~(\ref{smith:eq:TrapTc}) and (\ref{smith:eq:Muc}) reveals that the experimentally observed $T_c$ shift qualitatively mirrors the expected shift in $\mu_c$.
This similarity can be explained as follows: (i) The interaction shift of $\mu_c$ affects the density everywhere in the trap, (ii) Outside the small critical region the local density shift is simply proportional to the local $\mu$ shift. (iii) The $N_c$ (or $T_c$) shift from the non-critical region is thus proportional to the $\mu_c$ shift and greatly outweighs the contribution from the $n_c$ shift within the critical region.
So the beyond-MF $T_c$ shift observed in a trapped gas is directly related to the quadratic beyond-MF $\mu_c$ shift rather than the linear $n_c$ shift. We are thus still lacking a direct measurement of the historically most debated $n_c$ shift.  The achievement of homogeneously trapped gases has now brought such a measurement within reach.

\section{Critical Exponents of an Interacting Bose Gas}

Having discussed the location of the critical point we now briefly discuss the critical behaviour around that point.

The smallness of the critical region for a harmonically trapped gas places limitations on the critical behaviour that can be measured in these systems. This issue can be partly overcome by performing local measurements; such an approach was put to beautiful effect by Donner $et$ $al$ \cite{Donner:2007} who used an RF out-coupling technique to measure the divergence of the correlation length close to $T_c$ and obtained the critical exponent $\nu=0.67 \pm 0.13$, in agreement with the expected beyond-MF exponent for the 3D XY model.

\begin{figure} [t]
\centering
\includegraphics[width=1\columnwidth]{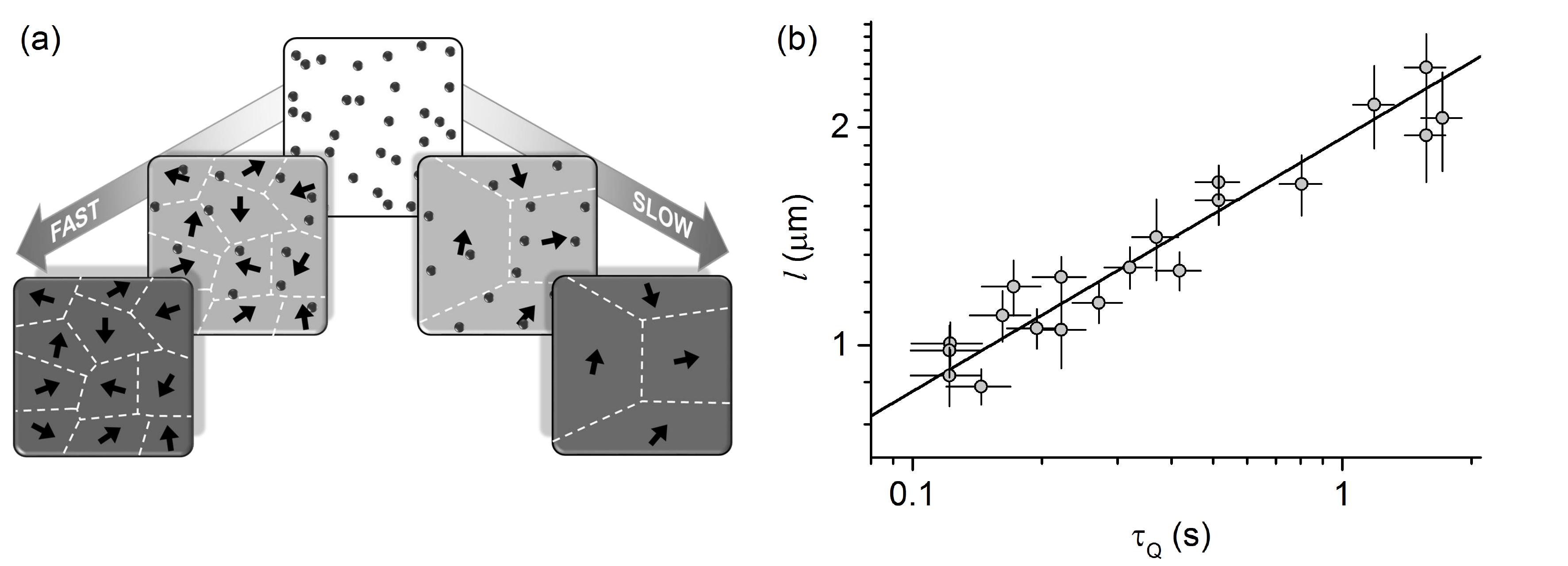}
\caption{ Critical exponents of the interacting BEC transition.
(a) The average size of the domains formed on crossing the critical point depends on the cooling rate.
(b) Scaling of domain size ($ \propto l$) with quench time $\tau_Q$. The data shows the expected Kibble-Zurek scaling $l \sim t_Q^b$ with $b=0.35(4)$ (solid line). This is in agreement with the beyond-MF prediction of $b\approx1/3$, corresponding to $\nu\approx 0.67$ and $z=3/2$.
Figures adapted from \cite{Navon:2015}.}
\label{smith:fig:KZ}
\end{figure}

The advent of homogeneous atom traps has opened up many more possibilities for the measurement of critical phenomena.  The first of these measurements for a 3D atomic Bose gas focused on the dynamics of spontaneous symmetry breaking at the BEC transition \cite{Navon:2015}. As we approach a second order transition the relaxation time ($\tau$), required to establish the diverging correlation length, also diverges. This divergence is described by $\tau \sim \xi^z \sim |t|^{-\nu z}$ where $z$ is the dynamical critical exponent. The consequence of this diverging $\tau$ is that as the transition is approached at any finite rate the system cannot adiabatically follow the diverging equilibrium $\xi$.  As a result the transition is crossed at a finite value of $\xi$, leading to the formation of domains  with independent choices of the symmetry breaking order parameter as shown in Fig.~\ref{smith:fig:KZ}(a). In a Bose gas this results in domains each of which is characterised by a wavefunction with a different phase. Kibble-Zurek theory describes how the length-scale $l$ associated with these domains scales with the speed of the quench, and predicts that $l \sim \tau_Q^b$ where $\tau_Q$ defines the quench rate across the transition via $\dot{t}=1/\tau_Q$ and $b=\nu/(1+\nu z)$.  Beyond-MF dynamical critical theory \cite{Hohenberg:1977} predicts $z=3/2$; combining this with the established $\nu=0.67$ gives $b\approx1/3$. Measurements on a homogeneous Bose gas of $^{87}$Rb atoms with $a/\lambda=0.008$ shown Fig. \ref{smith:fig:KZ} give $b=0.35 \pm 0.04$ in agreement with this expected scaling. This work not only confirmed the expected critical exponents for the BEC transition but also provided one of the first quantitative tests for Kibble-Zurek theory.

\section{Conclusion and Outlook}

In this chapter we have explored the effects of weak repulsive interactions on the condensation of atomic Bose gases. We have seen that the consequences of interactions depend strongly on whether we have a homogeneous system or one that is harmonically trapped.
In general, the presence of a harmonic trapping potential tends to magnify the effect of mean-field interactions while making the more interesting beyond-MF critical behaviour harder to observe. This makes the recent advances in studying Bose gases in homogeneous potentials particularly exciting. These advances promise the ability to study, in greater depth than has so far been possible, many interesting effects of interactions on both the thermodynamics and dynamics of Bose-Einstein condensation.

We have focused in this chapter on weak interactions, but a very interesting open question is what happens in the other extreme when interactions are as strong as possible. This regime, known as the unitary regime, happens when the scattering length $a$ tends to infinity and thus ceases to be a relevant scale in the problem. At this point the behaviour of the gas should be universal - only depending on the density. Experimental studies in this unitary regime are more difficult due to the rapid 3-body losses that occur for large $a$, and only recently are any results beginning to emerge \cite{Rem:2013, Fletcher:2013, Makotyn:2014}.


\end{document}